\documentstyle[iopl12]{ioplppt} 
\begin{document}
\mbox{} 
\vspace{2cm}
\begin{center}
        {\bf On Matrix Product States for Periodic Boundary Conditions} \\[20mm]
\end{center}
\begin{center}
\normalsize
         Klaus Krebs\\
        {\it Universit\"{a}t Bonn,
        Physikalisches Institut \\ Nu\ss allee 12,
   D-53115 Bonn, Germany}
\end{center}
\vspace{1cm}
%
        \vspace{2.5cm}
{\bf Abstract:}
The possibility of a matrix product representation for eigenstates with energy and momentum zero of 
a general $m$--state quantum spin Hamiltonian with nearest neighbour interaction 
and periodic boundary condition is considered.
The quadratic algebra used for this representation is generated by $2m$ operators 
which fulfil $m^2$ quadratic relations and is endowed with a trace.
It is shown that {\em not} every eigenstate with energy and momentum zero
can be written as matrix product state.
An explicit counter--example is given. 
This is in contrast to the case of open boundary conditions where every zero energy eigenstate
can be written as a matrix product state using a Fock--like representation of the 
same quadratic algebra.
\normalsize
\thispagestyle{empty}
\mbox{}       
\setcounter{page}{0}
\newpage
In a previous paper \cite{ks} it was shown that every zero energy eigenstate of a general 
$m$--state quantum spin Hamiltonian with nearest neighbour interaction in the bulk 
and open boundary conditions (i.\ e.\ with single site terms at each boundary)
can be written with the help of a quadratic algebra.  
In this letter we ask for an analogous statement for periodic boundary conditions.
By giving a counter--example we show that such a statement does not hold for periodic 
boundary conditions.
Let us start with a summary of the previous paper \cite{ks}.
Consider a Hamiltonian of the following form:
\begin{equation}
H_{\mbox{\scriptsize op}} = h_1 + \sum_{j=1}^{L-1}h_{j,j+1} + h_L.
\label{Hopen}
\end{equation}
The bulk interaction term $h_{j,j+1}$ acts locally on the sites
$j$ and $j+1$ and is defined by
\begin{equation}
h_{j,j+1} = \sum_{\mu,\nu,\sigma,\tau= 1}^{m}
\gamma^{\mu\nu}_{\sigma\tau} E_j^{\sigma\mu} E_{j+1}^{\tau\nu}
\label{hambulk}
\end{equation}
where the $E^{\sigma\tau}$ are $m\times m$--matrices with entries
$(E^{\sigma\tau})_{\mu,\nu} = \delta_{\sigma,\mu} \delta_{\tau,\nu}$.
The boundary terms $h_1$ and $h_L$ act on the sites  1 and $L$, respectively, and have the form
\begin{equation}
h_1 =   \sum_{\mu,\sigma= 1}^{m} \alpha^{\mu}_{\sigma}  E_1^{\sigma\mu} \qquad
h_L =   \sum_{\mu,\sigma= 1}^{m} \beta^{\mu}_{\sigma}  E_L^{\sigma\mu}.
\label{hambound}
\end{equation}
Following the suggestion of \cite{hsp}, for each Hamiltonian of this 
type we introduce a quadratic algebra
generated by  $2m$ generators $D_1,\cdots ,D_m$, $X_1,\cdots ,X_m$ 
fulfilling the following $m^2$ quadratic relations determined by the coefficients of the bulk
interaction term (\ref{hambulk}):
\begin{equation}
\sum_{\mu,\nu = 1}^{m}\gamma^{\mu\nu}_{\sigma\tau}
D_{\mu}D_{\nu} = X_{\sigma}D_{\tau} - D_{\sigma}X_{\tau} \qquad
\sigma, \tau = 1,\cdots,m.
\label{qalg}
\end{equation}
By giving a representation it was shown in \cite{ks} that this algebra 
exits for every choice of the coefficients $\gamma^{\mu\nu}_{\sigma\tau}$ in (\ref{qalg}).
Furthermore we introduce a Fock--representation of this algebra.
We assume that there is an auxiliary vector space $\cal V$, where the generators 
$D_1,\cdots ,D_m$, $X_1,\cdots ,X_m$ act on, and states $| V \rangle$ and $\langle W |$ 
in $\cal V$ and its dual, respectively, such that the following relations hold:
\begin{equation}
\sum_{\mu=1}^{m} \alpha^{\mu}_{\sigma} \langle W | D_{\mu} = - \langle W | X_{\sigma} \qquad
\sum_{\mu=1}^{m} \beta^{\mu}_{\sigma} D_{\mu} | V \rangle =  X_{\sigma} | V \rangle
\qquad \sigma = 1,\cdots,m.
\label{fockrep}
\end{equation}
The theorem proved in \cite{ks} makes two statements.
The first one says that the state  $P$ defined by
\begin{equation}
P = \sum_{\tau_1, \tau_2, \cdots, \tau_L = 1}^{m} 
\langle W |D_{\tau_1} D_{\tau_2} \cdots D_{\tau_L} | V\rangle
u^{(1)}_{\tau_1} \otimes u^{(2)}_{\tau_2} \otimes \cdots \otimes u^{(L)}_{\tau_L},
\label{mpgopen}
\end{equation}
where $u^{(k)}_{\tau}$ ($\tau = 1,\cdots,m$ and $k=1,\cdots ,L$) denotes 
the basis of the vector space of the $k$th site,
is an eigenstate of $H_{\mbox{\scriptsize op}}$ (\ref{Hopen}) with energy zero, i.\ e.\ 
$H_{\mbox{\scriptsize op}} P = 0$.
This can be shown using only the relations (\ref{qalg}) and (\ref{fockrep}).
A state of the form (\ref{mpgopen}) is called a matrix product state.
The second statement says that for {\it every} zero energy eigenstate 
$P'$ of $H_{\mbox{\scriptsize op}}$ one can find a representation of the operators 
$D_\tau$, $X_\tau$ and vectors $\langle W|$, $| V \rangle$ 
such that $P'$ can be written in the form (\ref{mpgopen}).
We would like to stress that the bulk algebra (\ref{qalg}) exists for every choice of the coefficients 
$\gamma^{\mu\nu}_{\sigma\tau}$ whereas the existence of the Fock representation (\ref{fockrep}), 
more precisely of the vectors  $\langle W|$ and $| V \rangle$, depends on the 
existence of a zero energy eigenstate of $H_{\mbox{\scriptsize op}}$.

Now we turn to periodic boundary conditions and ask for statements analogous to that described above. 
We consider a Hamiltonian of the form
\begin{equation}
H_{\mbox{\scriptsize per}} = \sum_{j=1}^{L} h_{j,j+1} \label{Haper}
\end{equation}
with periodic boundary conditions where the bulk interaction term 
$h_{j,j+1}$ is again given by (\ref{hambulk}).
As in the open boundary case, to each Hamiltonian of the form (\ref{Haper}) we 
may again associate the quadratic algebra (\ref{qalg}) which is determined just by the 
coefficients of $h_{j,j+1}$.
On this algebra we introduce a {\it trace--like} function. 
By this we mean a linear number--valued function which is invariant under cyclic permutations,
i.\ e.\  for a trace--like function $tr$ and for any two elements $A$ and $B$ 
of the algebra the relation 
\begin{equation}
tr(AB) = tr(BA)
\label{trace}
\end{equation}
holds. 
If a non--trivial trace--like function $tr$ exists on the quadratic algebra (\ref{qalg}), 
the state $P_0$ defined by
\begin{equation}
P_0 = \sum_{\tau_1 \cdots \tau_L = 0}^{m-1} tr(D_{\tau_1} \cdots D_{\tau_L})
u^{(1)}_{\tau_1} \otimes u^{(2)}_{\tau_2} \otimes \cdots \otimes u^{(L)}_{\tau_L}
\label{P0}
\end{equation}
is an eigenstate of $H_{\mbox{\scriptsize per}}$ with energy and momentum zero (cf.\ \cite{ahr}). 
This corresponds to the first statement of the theorem of \cite{ks}.
To get the analogue of the second statement we ask the following question:
Given an eigenstate $P_0'$ of $H_{\mbox{\scriptsize per}}$ with energy and momentum zero,
we ask for the corresponding trace--like function $tr_{P_0'}$ such that $P_0'$ is obtained by (\ref{P0}).
As we will see below, such a trace--like function does not exist in every case. 
We will present an example of a Hamiltonian which has an eigenstate with energy and
momentum zero which cannot be written in the form (\ref{P0}).

Before we come to this counter--example we would like to add some remarks on 
the relations between trace--like functions on the algebra (\ref{qalg}) 
and traces of representations of (\ref{qalg}).
It is not clear that any trace--like function can be obtained as 
the trace of an appropriate representation.
Consider for example the algebra discussed in \cite{ahr} (section 3.2 in this paper).
This algebra can be decomposed into two sectors.
For each sector a representation with a well--defined and non--trivial trace is given in \cite{ahr}.
However, a representation which has a trace on both sectors at the same time 
is not known to exist whereas a trace--like function with this property can easily be defined.
Next we would like to mention that a trace--like function can be defined 
without using an appropriate representation with a trace on it at all.
For example consider the coset algebra of (\ref{qalg}),
which is obtained by taking $X_\tau=0$ for $\tau = 1, \cdots, m$,
without any additional structure.
If this coset algebra exists the coefficients of the independent monomials  
are the weights of zero energy eigenstates of the corresponding Hamiltonian with 
closed boundary conditions 
(i.\ e.\ the Hamiltonian (\ref{Hopen}) with $h_1 = h_L = 0$) \cite{adr}.
In some cases there are sectors of this algebra where the coefficients of the
independent monomials are invariant under cyclic permutations.
In these cases the approach of \cite{adr} also works for the corresponding models
with periodic boundary conditions (but only in the corresponding sectors)
and the mapping which assigns to each monomial its coefficients of 
the independent monomials defines a trace--like function.
This happens in the case of the three state diffusion model considered in \cite{mev}.
(It is interesting to note that only the sector with equal number of particles for all species
was treated in \cite{mev}.
One can show that on all other sectors of this algebra a trace--like function
does not exist.)
Another useful observation is that the relations (\ref{qalg}) are homogeneously quadratic. 
Therefore they do not lead to relations of monomials or trace--like functions 
of monomials of different lengths.
Hence, on each sector generated by all monomials of a given length $L$,
a trace--like function can by defined separately.
We will make use of this fact later. 

Now we give an example for a Hamiltonian which has an eigenstate with
energy and momentum zero which cannot be written in the form (\ref{P0})
since for this state an appropriate trace--like function on the algebra (\ref{qalg}) 
does not exist.
We consider the Hamiltonian $H_3$ which is defined as the Hamiltonian (\ref{Haper}) on 
a 3--site lattice with 3 states per site and the following coefficients of the bulk 
interaction term (\ref{hambulk}):
\begin{equation}
 \gamma^{12}_{21} =  \gamma^{21}_{12} = -\gamma^{31}_{13} = -\gamma^{32}_{23} =  
-\gamma^{12}_{12} = -\gamma^{21}_{21} =  \gamma^{31}_{31} =  \gamma^{32}_{32} =  \alpha 
\label{contra}
\end{equation}
All other coefficients are zero.
This Hamiltonian has the following conserved quantities:
\begin{equation}
N_{\tau} = \sum_{j=1}^{3} E^{\tau\tau}_j \qquad \tau = 1,2
\label{conserv}
\end{equation}
A straightforward but lengthy calculation shows that on the zero momentum sector 
all matrix elements of $H_3$ vanish.
Hence, every translational invariant state has energy zero.
We call this sector ${\cal U}_0$.
It´s dimension is 11.
We ask which of the states of ${\cal U}_0$ can be obtained with the help of a trace--like function
on the algebra (\ref{qalg}). 
Therefore we have to look for the conditions which the quadratic relations (\ref{qalg}) 
and the cyclic invariance of the trace (\ref{trace}) put onto the values of
any trace--like functions.
If $tr$ is such a trace--like function these conditions have the form
\begin{eqnarray}
tr(X_{\tau_1} D_{\tau_2} D_{\tau_3}) - tr(X_{\tau_2} D_{\tau_3} D_{\tau_1})
&=& \sum_{\mu,\nu} \gamma^{\mu\nu}_{\tau_1 \tau_2} tr(D_{\mu} D_{\nu} D_{\tau_3}) \label{nec1}\\
tr(X_{\tau_1} X_{\tau_2} D_{\tau_3}) - tr(X_{\tau_3} X_{\tau_1} D_{\tau_2})
&=& \sum_{\mu,\nu} \gamma^{\mu\nu}_{\tau_2 \tau_3} tr(X_{\tau_1} D_{\mu} D_{\nu}) \label{nec2}
\end{eqnarray}
where $\tau_1,\tau_2,\tau_3 = 1,2,3$.
To simplify the notation for the following calculations we use the invariance under cyclic 
permutations to pushed the $X$--generators in the argument of $tr$ to the first positions
as done in (\ref{nec1}),(\ref{nec2}).
We consider (\ref{nec1}) and (\ref{nec2}) as a system of equations whose 
solutions are trace--like functions.
For each state in ${\cal U}_0$, which can be written in the form (\ref{P0}), 
there is a class of solutions of (\ref{nec1}) and (\ref{nec2}).
(A class of solutions consists of all trace--like functions whose values differ only on the 
monomials with one or more $X$--generators.) 
Our question is now, how many independent classes of solutions of (\ref{nec1}) and (\ref{nec2}) 
can be found.
It turns out that only 10 independent classes of solutions exist corresponding to 10 states 
of the form (\ref{P0}).
Thus, one state out of 11 remains which cannot be written in this form.
Let us now study solutions in detail.

The sector ${\cal U}_0$ can be divided up into the sub--sector of 
symmetric states (states which are 
invariant not only under cyclic permutation but under arbitrary permutations) 
and its orthogonal complement.
The symmetric sector has dimension 10.
All symmetric states can be written with the help of a trace--like function 
on the algebra (\ref{qalg}).
To see this we consider one--dimensional representations, i.\ e.\ we choose 
the generators to be numbers and the trace--like function to be the ordinary 
product of numbers.
In this case the quadratic relations (\ref{qalg}) with the coefficients of 
(\ref{contra}) reduce to the equations
\begin{eqnarray}
0              &=& d_1 x_2 - d_2 x_1  \nonumber \\
\alpha d_1 d_3 &=& d_1 x_3 - d_3 x_1  \label{1-d-rep} \\
\alpha d_2 d_3 &=& d_2 x_3 - d_3 x_2  \nonumber
\end{eqnarray}
where we have replaced the capital letters in (\ref{qalg}) by small ones.
It is easy to check that for each choice of $d_1$, $d_2$ and $d_3$ one
can find numbers $x_1$, $x_2$ and $x_3$ such that the relations (\ref{1-d-rep}) are fulfilled.
Therefore the values of $d_1$, $d_2$ and $d_3$ can be chosen arbitrarily.
For one--dimensional representations the coefficients of the state (\ref{P0}) 
are just the independent monomials of degree 3 in the 3 variables $d_1$, $d_2$ and $d_3$.
There are 10 of such monomials allowing as many independent states of the form (\ref{P0}).
Taking into account that a linear combination of trace--like functions is again 
a trace--like function every symmetric state can be written in the form (\ref{P0}). 
The orthogonal complement of the symmetric sector is generated by the single state $\varphi$
defined by
\begin{equation}
\varphi = \sum_{\sigma \in S_3} \mbox{sign}(\sigma)\;
u^{(1)}_{\sigma(1)} \otimes u^{(2)}_{\sigma(2)} \otimes u^{(3)}_{\sigma(3)}
\label{nogostate}
\end{equation}
where $S_3$ is the group of permutations of three objects. 
This state {\it cannot} be written with the help of a
trace--like function.
To show this we attempt to construct a function $tr_{\varphi}^*$ which fulfils all 
properties of a trace--like function on the algebra (\ref{qalg}) with the coefficients (\ref{contra})
and allows us to write the state $\varphi$ in the form (\ref{P0}).
(The star indicates that it is not yet clear whether $tr_{\varphi}^*$ can be extended to 
a trace--like function on all monomials of length 3.)
As pointed out above it is sufficient to define $tr_{\varphi}^*$ only for monomials of length 3.
For all further calculations the function $tr_{\varphi}^*$ is assumed to be invariant under cyclic 
permutations.
The values of $tr_{\varphi}^*$ for monomials of length 3 containing only $D$--generators are
uniquely determined by the requirement that $tr_{\varphi}^*$ allows a representation
of $\varphi$ in the form (\ref{P0}), i.\ e.\ 
\begin{equation}
tr_{\varphi}^*(D_{\sigma(1)} D_{\sigma(2)} D_{\sigma(3)}) = \mbox{sign}(\sigma) \qquad \sigma \in S_3.
\label{DDD}
\end{equation}
The values of $tr_{\varphi}^*$ on all other monomials containing only $D$--generators 
have to be zero.
The next step is to determine the values of $tr_{\varphi}^*$ on the monomials with 
one $X$--generator.
These values are constrained by the necessary conditions (\ref{nec1}) and (\ref{nec2}) 
which immediately result from 
the quadratic relations (\ref{qalg}) and the invariance of the trace--like function under 
cyclic permutations (\ref{trace}) and thus hold for any trace--like function $tr$.
We claim that $tr_{\varphi}^*$ is a trace--like function and use the necessary 
condition (\ref{nec1}) to determine the values of $tr_{\varphi}^*$ for monomials with one $X$--generator.
Note that on the left hand side of (\ref{nec1}) only differences of traces--like functions 
of monomials appear whose indices differ through a cyclic permutation. 
Hence, if $tr_{\varphi}^*(X_{\tau_1} D_{\tau_2} D_{\tau_3}) = t_{\tau_1\tau_2\tau_3}$
is a solution of (\ref{nec1}) for given values of  
$tr_{\varphi}^*(D_{\tau_1} D_{\tau_2} D_{\tau_3})$ then 
$tr_{\varphi}^*(X_{\tau_1} D_{\tau_2} D_{\tau_3}) = t_{\tau_1\tau_2\tau_3} + w_{\tau_1\tau_2\tau_3}$,
where the $w_{\tau_1\tau_2\tau_3}$ are arbitrary numbers with 
$w_{\tau_1\tau_2\tau_3} = w_{\tau_2\tau_3\tau_1}$,
is again a solution of (\ref{nec1}). 
Making use of this ambiguity, the values $tr_{\varphi}^*(X_{\tau_1} D_{\tau_2} D_{\tau_3})$
determined by (\ref{nec1}) can be written in the form
\begin{eqnarray}
tr_{\varphi}^*(X_1 D_2 D_3) &= w_{123} + \alpha \qquad
&tr_{\varphi}^*(X_1 D_3 D_2) = w_{132} - \alpha \nonumber \\ 
tr_{\varphi}^*(X_2 D_3 D_1) &= w_{123} + \alpha \qquad
&tr_{\varphi}^*(X_2 D_1 D_3) = w_{132} - \alpha \label{XDD} \\
tr_{\varphi}^*(X_3 D_1 D_2) &= w_{123} \qquad
&tr_{\varphi}^*(X_3 D_2 D_1) = w_{132} \nonumber 
\end{eqnarray} 
where $w_{123}$ and $w_{132}$ are arbitrary constants.
Up to these constants, the values of  $tr_{\varphi}^*$ in (\ref{XDD}) are uniquely determined
by (\ref{nec1}).
The values of $tr_{\varphi}^*$ on all other monomials with one $X$--generator
do not appear in further calculations. 
It remains to check whether condition (\ref{nec2}) can be fulfilled by the 
function  $tr_{\varphi}^*$ defined so far. 
Therefore we define the following function:
\begin{equation}
\fl
Z_{\tau_1 \tau_2 \tau_3}[tr] = \sum_{\mu\nu}\left[
\gamma^{\mu\nu}_{\tau_1 \tau_2} tr(X_{\tau_3} D_{\mu} D_{\nu}) +
\gamma^{\mu\nu}_{\tau_2 \tau_3} tr(X_{\tau_1} D_{\mu} D_{\nu}) +
\gamma^{\mu\nu}_{\tau_3 \tau_1} tr(X_{\tau_2} D_{\mu} D_{\nu}) \right]. 
\end{equation}
From (\ref{nec2}) we find $Z_{\tau_1 \tau_2 \tau_3}[tr] = 0$
as a necessary condition for $tr$ being a trace--like function.
Taking the values of $tr_{\varphi}^*$ defined in (\ref{XDD}) we find 
\begin{equation}
Z_{012}[tr_{\varphi}^*] = 2 \alpha^2  \neq 0. \label{nogo}
\end{equation}
Hence, the function $tr_{\varphi}^*$ defined above is {\it not} a trace--like function 
on the algebra (\ref{qalg}).
We would like to stress that there was no ambiguity in the definition of $tr_{\varphi}^*$ so far
up to the constants $w_{123}$ and $w_{132}$ which in turn do not appear in (\ref{nogo}).
Therefore a trace--like function on the quadratic algebra (\ref{qalg}) whose values on the 
monomials without $X$--generators are determined by (\ref{DDD}) does not exist at all.
Hence, the state $\varphi$ cannot be written in the form (\ref{P0})
with the help of an appropriate trace--like function.
This is what we wanted to show.
\newpage
\noindent
{\bf Acknowledgements}\\
I would like to thank Vladimir Rittenberg for many helpful discussions and for never--ending motivation. 
I would also like to thank  Ulrich Bilstein and Silvio Dahmen for critically reading the manuscript
and many helpful comments.
This work was supported by the TMR Network Contract FMRX-CT96-0012 of the European Commission.
\\[1cm]

\end{document}